\documentclass[a4paper]{IEEEtran}

\ifCLASSINFOpdf
   \usepackage[pdftex]{graphicx}
\else
   \usepackage[dvips]{graphicx}
\fi

\ifCLASSOPTIONcompsoc
  \usepackage[caption=false,font=normalsize,labelfont=sf,textfont=sf]{subfig}
\else
  \usepackage[caption=false,font=footnotesize]{subfig}
\fi

\usepackage{footnote}
\usepackage{nopageno}
\usepackage{float}
\usepackage{url}
\usepackage{amssymb,amsmath,amsfonts}
\usepackage{graphicx}
\usepackage{mathptmx}  
\usepackage{epstopdf}
\usepackage{textcomp}
\usepackage{xcolor}
\usepackage[absolute,showboxes]{textpos}
\usepackage{multirow}
\usepackage{multicol}
\usepackage{array} 
\usepackage{cite}
\usepackage{algorithm,algorithmic}
\usepackage{amsmath}
\usepackage[cmintegrals]{newtxmath}
\usepackage[utf8]{inputenc}
\usepackage[T5]{fontenc}
\usepackage{tikz}
\usetikzlibrary{positioning}
\usepackage{color,soul}

\begin{document}
\pagenumbering{gobble}

\title{$Q$-DATA: Enhanced Traffic Flow Monitoring in Software-Defined Networks applying $Q$-learning}
\author{
	\IEEEauthorblockN{Trung V. Phan\IEEEauthorrefmark{2}, Syed Tasnimul Islam\IEEEauthorrefmark{2}, Tri Gia Nguyen\IEEEauthorrefmark{4} and Thomas Bauschert\IEEEauthorrefmark{2}}\\
    \IEEEauthorblockA{\IEEEauthorrefmark{2}Chair of Communication Networks, Technische Universit{\"a}t Chemnitz,  09126 Chemnitz, Germany}\\
    \IEEEauthorblockA{\IEEEauthorrefmark{4}Faculty of Information Technology, Duy Tan University, Danang 50206, Vietnam}\\
    \IEEEauthorblockA {Email: trung.phan-van | thomas.bauschert@etit.tu-chemnitz.de}\\
}

\markboth{ACCEPTED for publication in IEEE CNSM conference 2019}%
{Trung~V.~Phan \MakeLowercase{\textit{et al.}}: $Q$-DATA: Enhanced Traffic Flow Monitoring in Software-Defined Networks applying $Q$-learning}

\maketitle

\begin{abstract}
Software-Defined Networking (SDN) introduces a centralized network control and management by separating the data plane from the control plane which facilitates traffic flow monitoring, security analysis and policy formulation. However, it is challenging to choose a proper degree of traffic flow handling granularity while proactively protecting forwarding devices from getting overloaded. In this paper, we propose a novel traffic flow matching control framework called $Q$-DATA that applies reinforcement learning in order to enhance the traffic flow monitoring performance in SDN based networks and prevent traffic forwarding performance degradation. We first describe and analyse an SDN-based traffic flow matching control system that applies a reinforcement learning approach based on $Q$-learning algorithm in order to maximize the traffic flow granularity. It also considers the forwarding performance status of the SDN switches derived from a Support Vector Machine based algorithm. Next, we outline the $Q$-DATA framework that incorporates the optimal traffic flow matching policy derived from the traffic flow matching control system to efficiently provide the most detailed traffic flow information that other mechanisms require. Our novel approach is realized as a REST SDN application and evaluated in an SDN environment. Through comprehensive experiments, the results show that---compared to the default behavior of common SDN controllers and to our previous DATA mechanism---the new $Q$-DATA framework yields a remarkable improvement in terms of traffic forwarding performance degradation protection of SDN switches while still providing the most detailed traffic flow information on demand.
\end{abstract}
\IEEEpeerreviewmaketitle

\section{Introduction}
Software Defined Networking (SDN) is a new networking concept, which provides enormous capabilities for dynamic network traffic control and management \cite{SurveySDN}. By detaching the control plane from the data plane, it removes some restrictions of legacy networks. A centralized entity called SDN controller has a global network view that allows for a policy-based traffic management and a faster and more dynamic response to network state and traffic variations \cite{SurveySDN}.

Although there are numerous available mechanisms for traffic analysis, traffic flow management and resilience \cite{TEinSDN} in SDN based networks, some significant challenges still remain to be addressed \cite{ChallengesforTEinSDN}. In particular, adapting the granularity of traffic forwarding while protecting forwarding devices from an overflow situation is one critical issue. As most current traffic management approaches rely on the default flow matching strategies of the available SDN controllers, it is difficult to perform traffic forwarding with variable granularity. For example, the \emph{Open Network Operating System} (ONOS) \cite{onos} and \emph{OpenDaylight} (ODL) \cite{odl} SDN controllers, by default, apply \emph{Reactive Forwarding} based on layer 2 information, which uses the \emph{MAC address} for flow matching only. Therefore, an incoming packet is matched to a flow entry by just using its layer 2 destination address. However, security and traffic monitoring mechanisms require traffic flow handling based on layer 3 and layer 4 information. A flow matching scheme that uses MAC and IP (and maybe also TCP/UDP) header fields requires much expensive TCAM memory for storing the respective flow rules \cite{SlowTCAM} and in case the incoming traffic flow pattern is highly dynamic, this might lead to a significant degradation of the traffic forwarding performance in the data plane. Besides, the control plane might be affected because of a large number of \textit{packet\_in} messages \cite{FloodGuard}.

In this paper, we propose a reinforcement learning based traffic flow matching control framework called $Q$-DATA, to enhance the performance of traffic flow monitoring in SDN based networks and proactively prevent flow-table overflow in SDN switches. We first describe a traffic flow matching control mechanism that applies a reinforcement learning based approach ($Q$-learning) for optimizing the traffic flow granularity in the data plane. It also considers the forwarding performance status of SDN switches derived by a Support Vector Machine algorithm. Next, we outline the $Q$-DATA framework that incorporates the optimal traffic flow matching policy derived from a $Q$-learning based Traffic Flow Matching Policy Creation module to efficiently provide detailed traffic flow information that other mechanisms, e.g., for traffic engineering, traffic monitoring, and intrusion detection, require. In particular, a Support Vector Machine algorithm is utilized to simultaneously analyse the current network traffic and predict the SDN switch performance degradation. Based on the prediction result the $Q$-learning based Traffic Flow Matching Policy Creation module issues an optimum action on changing the traffic flow matching scheme. Note that this proposal partially inherits\footnote{We leverage the use of the Support Vector Machine based performance degradation prediction mechanism and the traffic flow matching scheme change based on destination hosts from our previous study \cite{DATA}.} our previous work \cite{DATA} which is explained later on.

The paper is structured as follows. Section \ref{RelatedWork} provides related work and our previous study. Section \ref{RLforFlowMatchingChange} presents our approach for maximizing the level of traffic flow granularity based on $Q$-learning. Section \ref{Proposal} explains the $Q$-DATA framework in detail. Our experiments and results are outlined in section \ref{Experiments} and section \ref{ResultAnalysis}, respectively. Section \ref{Conclution} provides a summary and outlines some ideas for future studies.

\section{Related Work}\label{RelatedWork}
\subsection{Existing Methods for Flow Rule Control and Management in Software Defined Networks}
Many studies already addressed issues related to flow rule installation and management in SDN switches - a topic that is of high interest in the SDN research community \cite{STAR,LowOverheadFlowHolding,ReducingFlowTables,SimultaneouslyReducingLatency,Minnie:2017,Stephens:2012,Luo:2014,Mimidis:2016}. There exist several approaches for controlling TCAM utilization with the primary target of flow rule compression or aggregation.

The authors in \cite{STAR} propose an online routing scheme that constrains flow-table resources in SDN switches. Similarly, in \cite{LowOverheadFlowHolding} the objective is to maximize the number of flow entries in the data plane considering the limited flow-table space in SDN switches. Nonetheless, these methods do not address the problem of protecting the network infrastructure when a sudden traffic increase is happening.

The studies in \cite{ReducingFlowTables,SimultaneouslyReducingLatency,Minnie:2017} deal with TCAM resource management. In particular, \cite{ReducingFlowTables} outlines a solution for flow-table size reduction based on three criteria including Consistency\footnote{All the flows must be allotted with the same actions after the reduction.}, Absoluteness\footnote{All the manually added rules must be executed in the highest priority.} and Accuracy\footnote{The statistics data must be accurate all the time.}. An incoming packet classification approach is presented in \cite{SimultaneouslyReducingLatency} which exploits the temporal locality of network traffic to predict the flow of incoming packets. If the prediction is correct the forwarding latency and power consumption can be reduced trough avoiding the full flow-table lookup process in the TCAM. Rifai et al. introduce a framework called MINNIE \cite{Minnie:2017} for flow-table compression using wildcard rules. Furthermore, the authors in \cite{Stephens:2012} argue that for storing simple packet forwarding rules based on MAC addresses or VLAN IDs cheap SRAM memory is sufficient, while more complex matching rules (with more matching fields) might require the use of fast but expensive TCAM memory. Considering this, the amount of TCAM memory in SDN switches can be significantly reduced. The solution outlined in \cite{Luo:2014} applies the concept of flow rule aggregation by restructuring the matching fields. By that, the number of flow rules can be significantly reduced. Another approach for dynamic flow matching is proposed in \cite{Mimidis:2016} where a flow matching policy considering the DSCP values of different traffic types is applied.

The mentioned solutions only focus on flow-table size reduction and on enhancing the data plane forwarding performance. Contrary to our solution they do not consider the possibility of adaptively changing the traffic flow matching scheme depending on the current network state and the level of detail of traffic flow information that other mechanisms, for e.g., traffic engineering and monitoring, require.

\subsection{Destination-aware Adaptive Traffic Flow Rule Aggregation}\label{OurPreviousWork}
In our previous work, we proposed a destination-aware adaptive traffic flow rule aggregation solution named DATA \cite{DATA} for adapting the number of flow entries in SDN switches according to the level of detail of traffic flow information that other mechanisms require and at the same time preventing SDN switch performance degradation.

We analyzed common SDN flow matching strategies of the ONOS \cite{onos} and ODL \cite{odl} SDN controllers and their implications. We denoted the MAC Matching Only Scheme as \textit{MMOS} strategy and the Full Matching Scheme as \textit{FMS} strategy. Using MMOS the ability to track and monitor network traffic for security or forensic analysis is limited, whereas applying the FMS strategy can result in significant degradation of the forwarding performance or even to an SDN switch outage in case the maximum number of flow entries is reached. To solve this problem, we applied a 2-dimensional Support Vector Machine (SVM) algorithm \cite{SVMIntro} to anticipate the switch performance degradation well before it occurs and to trigger the flow matching scheme change in time. After analyzing the SDN switch performance and if a potential forwarding performance degradation is figured out, the Analyzer applies Algorithm \ref{FMStoMMOS} (see appendix section) to find some destination hosts whose associated flows are most critical regarding the forwarding performance of the SDN switch, i.e., have the most flow entries. Afterwards the Analyzer co-operates with the built-in forwarding application of the SDN controller to conduct the traffic flow matching scheme changes\footnote{In order to perform a traffic flow matching scheme change for a destination host, the built-in forwarding application firstly deletes all flow entries related to the destination host in the switch, then it installs a flow entry with a new match field combination in the switch.} for these destination hosts. These actions can be either to change to MMOS if a sudden increase or an overflow of the flow-table space in an SDN switch is expected or to return to FMS in case there is no overflow risk (see Algorithm \ref{MMOStoFMS} in appendix section).

Our DATA approach outperforms legacy flow rule matching schemes in terms of the number of flow entries in the SDN switches, the average $packet\_in$ rate in the SDN control plane and the number of errors and exceptions.

Although, the DATA method has many advantages in comparison to legacy approaches, some issues still should be addressed for further improvement - e.g., the limited number of only two flow matching schemes (MMOS and FMS) and the lack of feedback about the impact of the respective flow matching scheme on the network performance. Therefore, in this paper, we propose a novel traffic flow matching control mechanism that can flexibly switch between many different flow matching schemes based on the current network state. The novel scheme provides a much higher level of detail of traffic flow information even in case of high traffic load, while effectively preventing flow-table overflow and degradation of the data plane forwarding performance.

\begin{figure}
\centering
\includegraphics[width=0.46\textwidth]{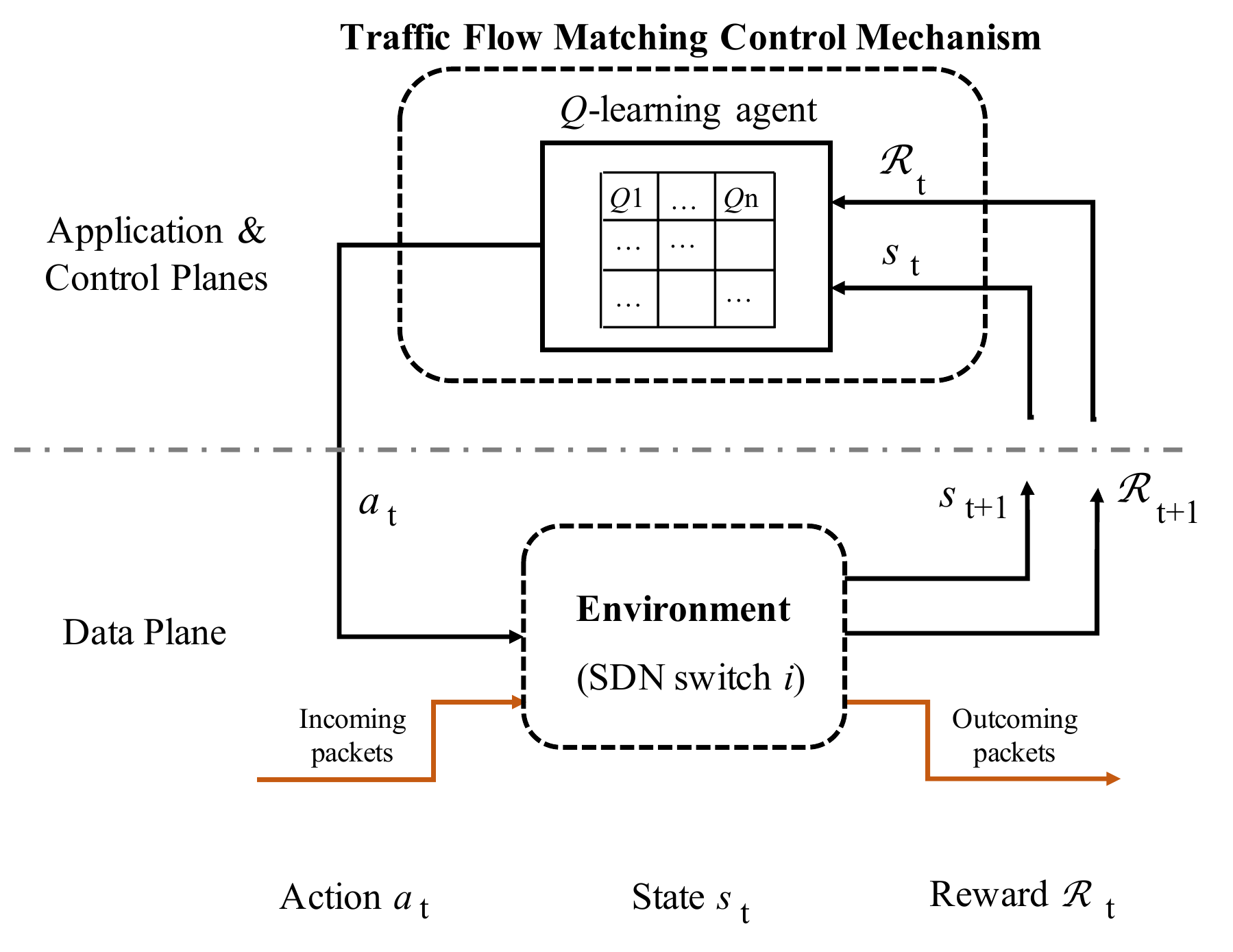}
\caption{Reinforcement learning-based model of a traffic flow matching control system in SDN based networks}
\label{fig:Q-LearningModel}
\end{figure}

\section{Maximizing the traffic flow granularity applying a $Q$-learning algorithm}\label{RLforFlowMatchingChange}
Fig. \ref{fig:Q-LearningModel} shows a traffic flow matching control mechanism based on reinforcement learning. The traffic flow matching control mechanism is realized as an SDN application and the environment is represented by the devices in the data plane, i.e., the SDN switches. In the following, we have a look at a single SDN switch $i$ representing the environment and investigate the traffic flow matching control mechanism. We assume that a state $s_{t}$ of the SDN switch $i$ at a time $t$ is represented by a tuple including the total number of current flow entries ($f_{i}$) and the number of flow entry changes ($\Delta f_{i}$) between two consecutive observations; the \textit{long-term} goal of the control system is to maximize the traffic flow granularity in state $s_{t}$ of SDN switch $i$ while protecting the switch from forwarding performance degradation.

Regarding the system operation, in a given state $s_{t}$ the control mechanism initiates an action $a_{t}$ to change the traffic flow matching scheme in SDN switch $i$. Afterwards, a new state $s_{t+1}$ is observed and a reward $\mathcal{R}_{t}$ is calculated as soon as the change of the traffic flow matching scheme is executed. Then the next action $a_{t+1}$ is applied to the environment in order to achieve the long-term goal. The traffic flow matching control mechanism based on reinforcement learning operates via agent-environment interaction and can be modeled as a Markov Decision Process (MDP) \cite{RLIntro}. In the following, the MDP model is outlined in detail.

\subsubsection{State Space} The state space of SDN switch $i$ is defined as follows:
\begin{equation}\label{StateSpace}
\mathcal{S}_{i} \triangleq \{\left(f_{i},\Delta f_{i}\right): 0 \textcolor{red}{<} f_{i}\leq f_{cap_{i}}; -f_{cap_{i}} \leq \Delta f_{i} \leq f_{cap_{i}}\},
\end{equation}
where $f_{i}$ is the current total number of flow entries in switch $i$, $\Delta f_{i}$ is the number of flow entry changes between two consecutive observations and $f_{cap_{i}}$ is the maximum number of flow entries in switch $i$. The state of SDN switch $i$ is defined as tuple $s$ = $\left(f_{i},\Delta f_{i}\right) \in \mathcal{S}_{i}$. In our previous study \cite{DATA} we already discussed the reasons for choosing the tuple ($f_{i}$,$\Delta f_{i}$) as the representative for the state of an SDN switch. The reasons can be summarized as follows: the effort for flow entry searching and matching in an SDN switch is proportional to the number of matching fields and an SDN switch has a maximum capacity ($f_{cap}$) for storing the flow entries. Accordingly, the change of the number of flow entries indicates the control plane load (wrt. \textit{of\_mod} and \textit{of\_removed} messages sent between SDN controller and switch) affecting both the SDN switch and the controller performance.

\subsubsection{Action Space} $\mathcal{F} = \{g_{1}, g_{2}, ..., g_{m}\}$ denotes a list of all feasible match field combinations, e.g., $g_{m}=$ <"\textit{matchTcpUdpPorts}", "\textit{matchIpv4Address}", "\textit{matchVlanId}",...> in case of the ONOS controller \cite{onos}. The action space for changing the traffic flow matching scheme in the SDN switch $i$ is defined by
\begin{equation}\label{ActionSpace}
\mathcal{A}_{i} \triangleq \{a:a \in \mathcal{F}\},
\end{equation}
where $a$ represents a traffic flow matching scheme change related to a destination host (as discussed in section \ref{OurPreviousWork}) in SDN switch $i$.

\subsubsection{Immediate Reward Function} 
On the one hand, whenever, through executing an action, the total number of current flow entries in the SDN switch $i$ reaches the limit $f_{cap_{i}}$ (which then leads to a performance degradation), the traffic flow matching control system should not get any reward for this action. On the other hand, the more matching fields a flow entry contains, the more detailed information is available for that flow. Hence, we determine the immediate reward as the average number of matching fields of all flow entries in the SDN switch $i$: 

\begin{equation}\label{RewardFunction}
	\mathcal{R}_{i}(s,a) = \left\{ \begin{array}{rl} \frac{\sum_{x=1}^{f_{i}}\Theta_{x}}{f_{i}}, & \quad 0 \textcolor{red}{<} f_{i} < f_{cap_{i}}, \\ 0, & \quad f_{i} = f_{cap_{i}}, 
	\end{array}\right.
\end{equation}
where $f_{i}$ is the current total number of flow entries in the switch $i$, $\Theta_{x}$ is an integer number representing the number of enabled match fields in flow entry $x$.

\subsubsection{Optimization Formulation} 
We define an optimization problem to acquire the optimal policy applicable in state $s$, denoted by $\pi^{*}(s)$, that maximizes the long-term reward, i.e., the traffic flow granularity in the SDN switch $i$ while protecting it from forwarding performance degradation. In particular, in state $s$, the agent issues an optimal action $a$ to get close to or reach the long-term reward. The MDP under consideration is finite and the state space $\mathcal{S}_{i}$ contains at maximum 2${f_{cap}}_{i}^{2}$ states. The optimization problem is formulated as follows: 
\begin{equation}\label{OptimizationProblem}
\begin{split}
\max_{\pi} \quad \{\mathfrak{R}(\pi)_{i}=\sum_{t=1}^{2{f_{cap}}_{i}^{2}} \mathbb{E} (\mathcal{R}_{i}(s_{t},\pi (s_{t}))): \mathcal{R}_{i} \in \mathbb{R}; s_{t} \in \mathcal{S}_{i}; \\ \pi (s_{t}) \in \mathcal{A}_{i}\} \\
\text{subject to} \quad SVM(s_{t})=\textit{Good}, \forall s_{t} \in \mathcal{S}_{i},
\end{split}
\end{equation}
where $\mathfrak{R}(\pi)_{i}$ is the cumulative reward for SDN switch $i$ under a policy $\pi$, $\mathcal{R}_{i}(s_{t},\pi (s_{t}))$ is the immediate reward associated with policy $\pi$ for a switch $i$ at iteration $t$, and SVM($s_{t}$) is the result of the Support Vector Machine algorithm predicting the forwarding performance of the SDN switch. A "\textit{Good}" result of the SVM algorithm means that the switch can handle the current number of flow entries without any forwarding performance problems.

In order to solve the optimization problem, we apply the $Q$-learning algorithm \cite{RLIntro} which uses a $Q$-table to represent all possible state-action pairs within the environment as shown in Fig. \ref{fig:Q-LearningModel}. The $Q$-learning agent can learn from its own decisions at each iteration, and the algorithm will converge to the optimal policy $\pi^{*}$ after a certain number of iterations \cite{RLIntro}. The expected return of state $s$ under policy $\pi$ is denoted as $\vartheta_{\pi}(s) : \mathcal{S}_{i} \longrightarrow \mathbb{R}$. It is expressed as follows: 
\begin{equation}\label{ValueFunction}
\begin{split}
\vartheta_{\pi}(s) = \mathbb{E}_{\pi} \left[ \sum_{t=0}^{2{f_{cap}}_{i}^{2}} \gamma \mathcal{R}_{i}(s_{t},a_{t}) | s_{t}=s \right] = \mathbb{E}_{\pi} [\mathcal{R}_{i}(s_{t},a_{t}) \\ + \gamma \vartheta_{\pi}(s_{t+1}) | s_{t}=s], \forall s \in \mathcal{S}_{i},
\end{split}
\end{equation}
where $\gamma\in$[0, 1) is a discount factor that indicates the importance of the long-term reward \cite{RLIntro}. The optimal policy $\pi^{*}$ in state $s$ represents an action $a$ that yields the maximum value of the expected return $\vartheta_{*}(s)$:
\begin{equation}\label{OptimalValueFunction}
\vartheta_{*}(s) = \max_{a} \left\{ \mathbb{E}_{\pi} \left[ \mathcal{R}_{i}(s_{t},a_{t}) + \gamma \vartheta_{\pi}(s_{t+1}) | s_{t}=s \right] \right\}, \forall s \in \mathcal{S}_{i}.
\end{equation}
Thus, for all state-action ($s$,$a$) pairs, the optimal $Q$-functions are
\begin{equation}\label{OptimalQ-Function}
\mathcal{Q}_{*}(s,a) \triangleq \mathcal{R}_{i}(s_{t},a_{t})+\gamma \mathbb{E}_{\pi} \left[ \vartheta_{\pi}(s_{t+1}) \right], \forall s \in \mathcal{S}_{i}.
\end{equation}
Hence $\vartheta_{*}(s)$ can be expressed as $\vartheta_{*}(s) = \max_{a} \left\{ \mathcal{Q}_{*}(s,a) \right\}$. By conducting different actions $a$ to the environment the optimal $Q$-function value, i.e., $\mathcal{Q}_{*}(s,a)$, for all state-action ($s$,$a$) pairs is figured out. In particular, the $Q$-function is updated at each iteration as follows:
\begin{equation}\label{UpdateQ-Function}
\begin{split}
\mathcal{Q}_{t+1}(s_{t},a_{t})=\mathcal{Q}_{t}(s_{t},a_{t})+\alpha [ \mathcal{R}_{i}(s_t,a_t)+\gamma\max_{a}\mathcal{Q}_{t}(s_{t+1},a) \\ - \mathcal{Q}_{t}(s_{t},a_{t})],
\end{split}
\end{equation}
where $s_{t}\in\mathcal{S}_{i}$, $a_{t}\in \mathcal{A}_{i}$. $\mathcal{Q}_{t}(s_{t},a_{t})$ is the $Q$-value for a state-action pair $(s_{t},a_{t})$, $\mathcal{R}_{i}(s_t,a_t)$ is the immediate reward for the SDN switch $i$ at an iteration $t$, $\gamma\in$[0, 1] is the discount factor and $\alpha\in$[0,1] is the learning rate. Moreover, to mitigate the exploration and exploitation dilemma that has direct impact on the convergence rate of any learning algorithms, the $\epsilon$-greedy algorithm \cite{RLIntro} is applied. Instead of always taking the best action according to the network state, the $Q$-learning agent will take some random actions, where the probability of a random decision is determined by the value of epsilon, $\epsilon$. In its learning phase, the $Q$-learning agent first of all arbitrarily initializes the $Q$-table for all state-action pairs and afterwards updates it by using Equation \ref{UpdateQ-Function}. Accordingly, the agent acquires a trained or converged $Q$-table.

In summary, the $Q$-learning agent generates the optimal policy $\pi^{*}(s)$ for a state $s$ representing an action $a$ that needs to be taken to maximize the value of the $\mathcal{Q}_{*}(s,a)$ function, i.e., $\pi^{*}(s)=\arg \max_{a}\mathcal{Q}_{*}(s,a)$. Algorithm \ref{AdaptiveFlowMatchingChangeAlgorithm} provides implementation details of the $Q$-learning algorithm.

\begin{algorithm}
\caption{Optimal traffic flow matching policy creation with $Q$-learning algorithm}
\label{AdaptiveFlowMatchingChangeAlgorithm}
\begin{algorithmic}[1]
\STATE \textbf{Inputs}: $\mathcal{F}$; for a state-action pair ($s$,$a$) $\forall s \in \mathcal{S}_{i}$, $a \in \mathcal{A}_{i}$, initialize a $Q$-table entry arbitrarily; initialize values of $\alpha$, $\gamma$, and $\epsilon$, respectively.
\LOOP
    \STATE Current state $s_{t}$.
    \STATE Execute action $a_{t}$ according to an exploratory policy ($\epsilon$).
    \STATE Obtain a new state $s_{t+1}$ and an immediate reward $\mathcal{R}_{i}$.
    \STATE Update the $Q$-table entry for $\mathcal{Q}$($s_{t}$,$a_{t}$) using Equation \ref{UpdateQ-Function}.
    \STATE Update $s_{t} \longleftarrow s_{t+1}$.
\ENDLOOP
\STATE \textbf{Outputs} $\pi^{*}(s) = \arg \max_{a} \mathcal{Q}_{*}(s,a)$.
\end{algorithmic} 
\end{algorithm}

\begin{figure}
\centering
\includegraphics[width=0.46\textwidth]{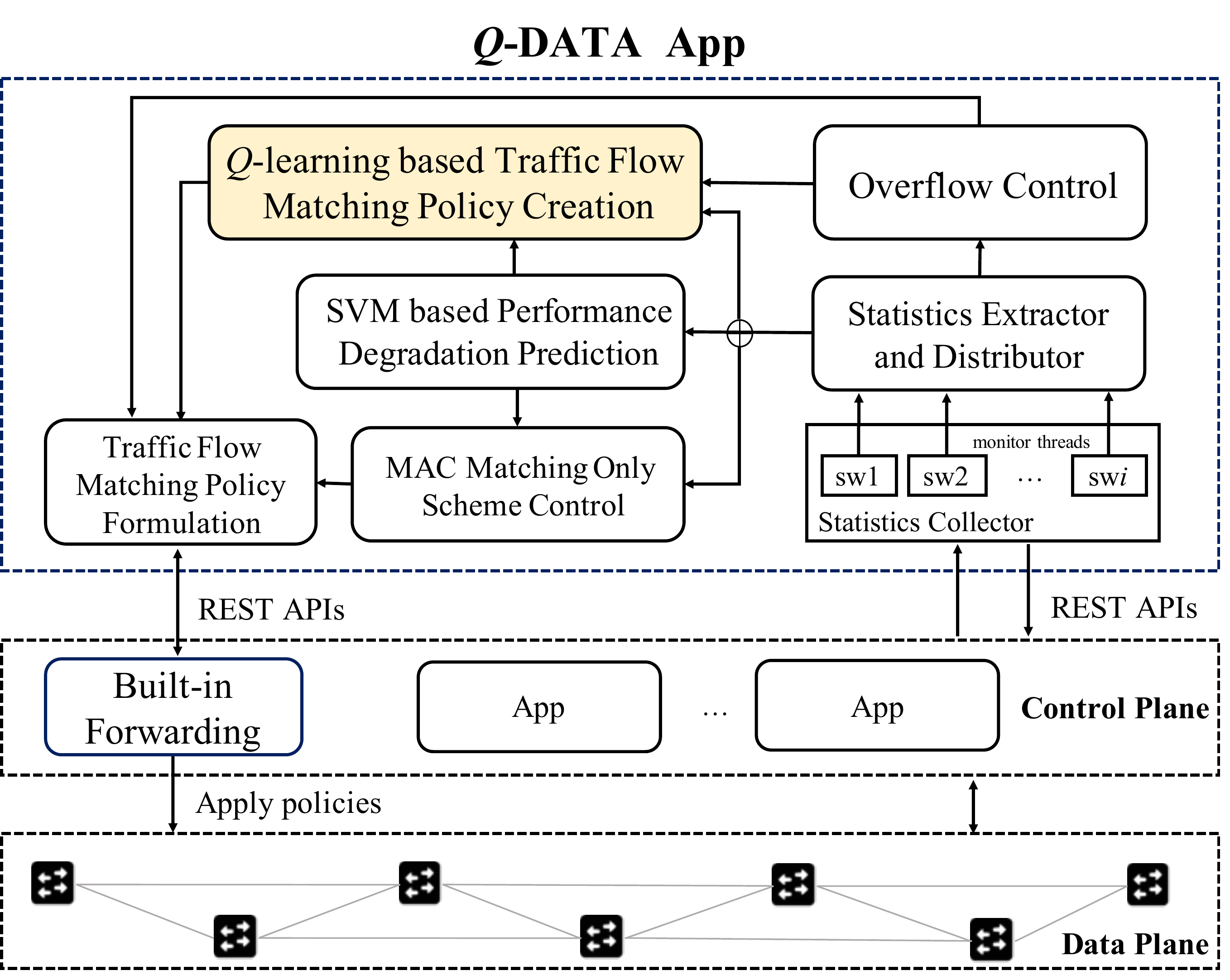}
\caption{$Q$-DATA framework architecture}
\label{fig:Q-DATA}
\end{figure}

\section{$Q$-DATA framework}\label{Proposal}
In this section, the design and operation of the $Q$-DATA framework for enhanced traffic flow monitoring and proactively prevention of forwarding performance degradation in SDN based networks is outlined.

\subsection{$Q$-DATA Framework Architecture}\label{SystemArchitectureDesign}
Fig. \ref{fig:Q-DATA} shows the $Q$-DATA framework architecture consisting of the Built-in Forwarding application located in the control plane and the REST $Q$-DATA application residing in the SDN application plane.

\subsubsection{Built-in forwarding application}
Most of the well-known SDN controllers \cite{onos,odl} provide basic forwarding functionality by running a built-in forwarding application to create flow rules which are then downloaded to the SDN switches. We propose to add a REST API interface to the built-in forwarding application to securely communicate with the $Q$-DATA application. Initially, the $Q$-DATA App instructs the built-in forwarding application to apply the Full Matching Scheme (FMS) strategy.

\subsubsection{$Q$-DATA App}
In $Q$-DATA a Statistics Collector periodically gets raw information about all traffic flows traversing the SDN switches from the SDN controller via the REST APIs \cite{onos,odl}. The collected statistical data of the SDN switch $i$ is forwarded to a Statistics Extractor and Distributor for extracting and distributing flow statistics information to other modules, i.e., the SVM based Performance Degradation Prediction module, the MAC Matching Only Scheme Control module, the Overflow Control module and the $Q$-learning based Traffic Flow Matching Policy Creation module. The SVM based Performance Degradation Prediction module is designed to anticipate the performance degradation of the SDN switch $i$ well before it occurs \cite{DATA} and to provide the prediction result to the $Q$-learning based Traffic Flow Matching Policy Creation module and the MAC Matching Only Scheme Control module. The Overflow Control module acts as an immediate reaction mechanism against a flow-table overflow situation, e.g., in case the network is under a Denial-of-Service attack. The MAC Matching Only Scheme Control module monitors and checks conditions for a traffic flow matching scheme change to FMS in the SDN switch $i$. The $Q$-learning based Traffic Flow Matching Policy Creation module relies as discussed above on a converged $Q$-table to choose the most appropriate traffic flow matching scheme for a given state of the SDN switch $i$. Finally, the Traffic Flow Matching Policy Formulation module formulates policies received from the Overflow Control, the MAC Matching Only Scheme Control and the $Q$-learning based Traffic Flow Matching Policy Creation modules and sends them to the Built-in Forwarding application for implementation in the SDN switch $i$.

\begin{figure}
\centering
\includegraphics[width=0.5\textwidth]{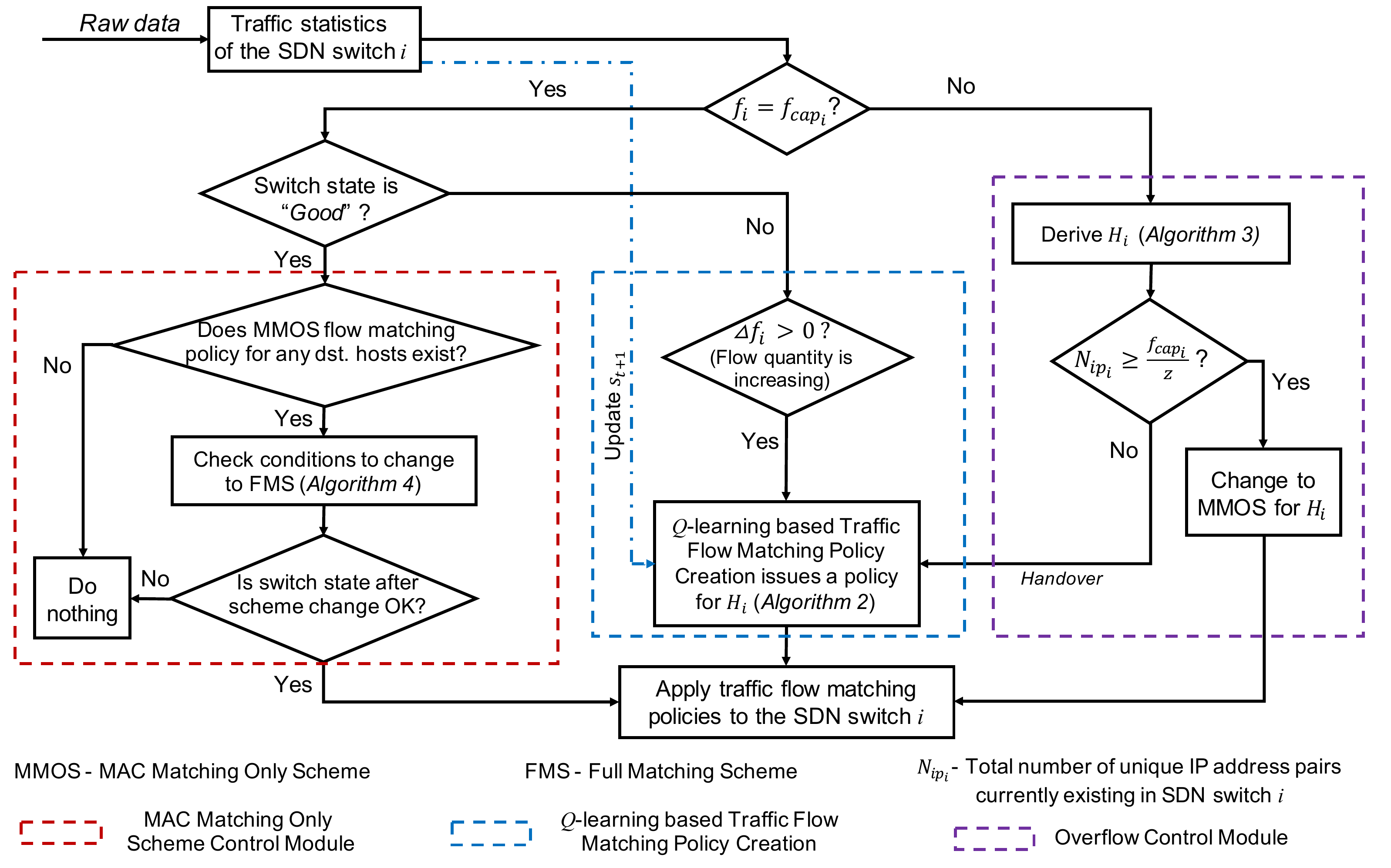}
\caption{Operational workflow of the $Q$-DATA framework}
\label{fig:FlowChart}
\end{figure}

\subsection{Operational Workflow}\label{OperationalSystemWorkflow}
Initially, the Statistics Collector sends a request to the SDN controller to ask for network topology information. Then, it launches a monitor thread $swi$ for each SDN switch $i$ - see Fig. \ref{fig:Q-DATA}. In regular time intervals (observation period), the monitor thread $swi$ gathers raw traffic flow statistics from the SDN switch $i$ and passes them to the Statistics Extractor and Distributor where the tuple ($f_{i}$,$\Delta f_{i}$)—the total number of current flows and the flow number changes—is determined. Afterwards, this data is forwarded to the Overflow Control module, the SVM based Performance Degradation Prediction module and the MAC Matching Only Scheme Control module.

Fig. \ref{fig:FlowChart} shows the detailed operational workflow of the $Q$-DATA framework. Firstly the Statistics Extractor and Distributor module compares $f_{i}$ to $f_{cap_{i}}$ and if the current total number of flow entries in a switch $i$ reaches its upper limit, then it is denoted as an overflow situation. In this case, the Overflow Control module has to find an appropriate traffic flow matching change policy for some destination hosts $H_{i}$ (derived from Algorithm \ref{FMStoMMOS}) which have most flow entries in switch $i$, so that the overflow situation is mitigated. We suppose that a source-destination node pair (having a unique pair of IP addresses), that transfers traffic through switch $i$, puts $z$ flow entries ($z \geq 1.0$) on average in the switch $i$ (e.g., either request or response flows). $N_{ip_{i}}$ denotes the total number of unique IP address pairs in the flow-table of the switch $i$. In case of a non-saturation attack the number of hosts (represented by their IP addresses\footnote{Initially the Built-in Forwarding application applies the FMS scheme, hence IP address information is available before an overflow problem appears in the switch $i$.}) sending traffic through the SDN switch $i$ is usually much less than the maximum number flow entries $f_{cap_{i}}$. Thus, if $N_{ip_{i}} \geq \frac{f_{cap_{i}}}{z}$, there are some destination hosts $H_{i}$ serving a lot of incoming requests from other hosts or being under saturation attacks (e.g., Denial-of-Service attacks). Hence, it is reasonable to match incoming traffic flows related to these destination hosts using only MAC address information. This avoids a sudden overflow situation of switch $i$. Otherwise, the Overflow Control module handovers to the $Q$-learning based Traffic Flow Matching Policy Creation module to issue an optimal traffic flow matching policy for the $H_{i}$ destination hosts via Algorithm \ref{Q-Learning-basedPolicyCreation}.

In case the current total number of flow entries $f_{i}$ is less than the switch's upper limit $f_{cap_{i}}$, the SVM based Performance Degradation Prediction module checks for a potential performance degradation of the SDN switch $i$ based on the tuple ($f_{i}$,$\Delta f_{i}$), and forwards its prediction result to the MAC Matching Only Scheme Control module and the $Q$-learning based Traffic Flow Matching Policy Creation module. If the switch state is predicted as \textit{Good}, the MAC Matching Only Scheme Control module checks whether there exists a MMOS flow matching policy for any of the destination hosts. If a MMOS flow matching policy is found, then Algorithm \ref{MMOStoFMS} is applied to check the conditions for a change to the FMS strategy. In case a possible performance degradation is detected for switch $i$, and if the total number of flow entries is increasing ($\Delta f_{i}$>0), then the $Q$-learning based Traffic Flow Matching Policy Creation module executes Algorithm \ref{Q-Learning-basedPolicyCreation} to apply the most appropriate traffic flow matching policy for destination hosts (derived from Algorithm \ref{FMStoMMOS}) in the switch $i$.

\begin{algorithm}
\caption{$Q$-learning based traffic flow matching policy creation for the SDN switch $i$}\label{Q-Learning-basedPolicyCreation}
\begin{algorithmic}
\STATE \textbf{Input}: A tuple $\left(f_i,\Delta f_i\right)$ at iteration $t$; $\mathcal{R}_{i}$; $H_{i}$.
\STATE \textbf{begin}
    \STATE \quad Utilize a converged $Q$-table from Algorithm \ref{AdaptiveFlowMatchingChangeAlgorithm}.
    \STATE \quad $s_{t} \gets$  $\left(f_i,\Delta f_i\right)$.
    \STATE \quad Drive optimal action $a_{t}$ = $\pi^{*}(s_{t})$ from the $Q$-table.
    \STATE \quad Apply $a_{t}$ for $H_{i}$ derived from Algorithm \ref{FMStoMMOS}.
    \STATE \quad Update $s_{t}$ $\gets$ $s_{t+1}$ \COMMENT{see Fig. \ref{fig:FlowChart}}.
    \STATE \quad Get reward $\mathcal{R}_{i}$ using Equation \ref{RewardFunction}.
    \STATE \quad Update $Q$-table using Equation \ref{UpdateQ-Function}.
    \STATE \textbf{end}
\end{algorithmic} 
\end{algorithm}

\section{Experiments}\label{Experiments}
\subsection{Example SDN Network Scenario}\label{DeploymentSetup}
In order to evaluate the performance of the $Q$-DATA framework, we leverage the MaxiNet framework \cite{MaxiNet} to emulate a simple SDN based network consisting of 3 Web servers (S1-S3) (using Apache Web server images) and 5 hosts (H1-H5) which are all connected to a single SDN switch (implemented as OpenvSwitch). The emulated SDN network runs within one Linux machine and is controlled by a remote ONOS SDN controller running on another physical machine. For ease of deployment, we place both the $Q$-DATA App and the ONOS SDN controller on the same Linux machine.

\subsection{Training $Q$-learning and SVM Algorithms}
Initially, for training the $Q$-learning agent we use the Hping3 tool \cite{hping3} installed in hosts (H1-H5) to randomly generate traffic between hosts and Web servers. The $Q$-learning agent depends on the collected data for making decisions about changing the traffic flow matching scheme, and for updating its $Q$-table accordingly. In particular, we set the $\epsilon$ value to 0.8 in order to have 80\% of random actions in a set of 9 match field combinations, and the state observation time is set to 10.0 seconds. For training the SVM algorithm, we apply the same traffic generation strategy as for the $Q$-learning agent training phase and initially apply the FMS scheme. Afterwards, we monitor any errors or exceptions indicating that the switch cannot handle new flow requests, and set $sign$ = -1 as a label for the associated tuple ($f_{i}$,$\Delta f_{i}$). Otherwise, we set $sign$ = +1. These labelled samples are then used for training the SVM algorithm.

We observe that the switch starts getting overflowed or cannot handle new flow rules if the current total number of flows is around 3000 ($f_{cap_{i}}$) \cite{SlowTCAM,DATA}. Setting the \textit{idle\_timeout} value (after which the flow entries are removed) to 10 seconds, the safety threshold for the packet rate the switch can handle is 300 packets per second assuming that each packet belongs to a different traffic flow rule (worst case assumption). Therefore, for traffic generation, we apply three levels: low load ($R1$=100), medium load ($R2$=200) and high load ($R3$=300).

\subsection{Experiment Setup}
We conduct several experiments with different flow matching strategies: MMOS only, FMS only, the novel $Q$-DATA framework (with $\epsilon=0.0$, $\epsilon=0.2$, $\epsilon=0.8$) and the DATA scheme \cite{DATA}. The built-in forwarding application of the ONOS SDN controller applies \emph{Reactive Forwarding}. 

In order to show the performance enhancement in traffic flow monitoring in SDN based networks with the $Q$-DATA framework, we implement a SOM-based IDS application (Self Organizing Map algorithm \cite{SOM}) to detect abnormal traffic on top of the ONOS controller. We consider some common attacks, which can make the SDN switch become overflowed, comprising TCP SYN flood \cite{DDoSinSDNSurvey}, Port scanning \cite{DDoSinSDNSurvey}, Low and Slow Denial-of-Service \cite{SlowTCAM}. The attack traffic is stemmed from hosts and it is directed to Web servers in our setup.

For the performance analysis, traffic from the 5 hosts towards the 3 servers is generated randomly with three different load levels ($R1$, $R2$, $R3$). During the experiments we trace the total number of flow entries in the SDN switch, the average number of \textit{packet\_in} messages per second to the ONOS controller, errors and exceptions in the ONOS controller, the frequency of traffic flow matching policy changes, the CPU utilization of the controller machine and the attack detection performance of the SOM-based IDS.

\section{Results}\label{ResultAnalysis}

\subsection{Network related Performance Results}
\subsubsection{Total number of traffic flow entries in the SDN switch}
\begin{figure*}
  \begin{minipage}{\linewidth}
  \centering
  \subfloat[]{
  \includegraphics[width=.3\linewidth]{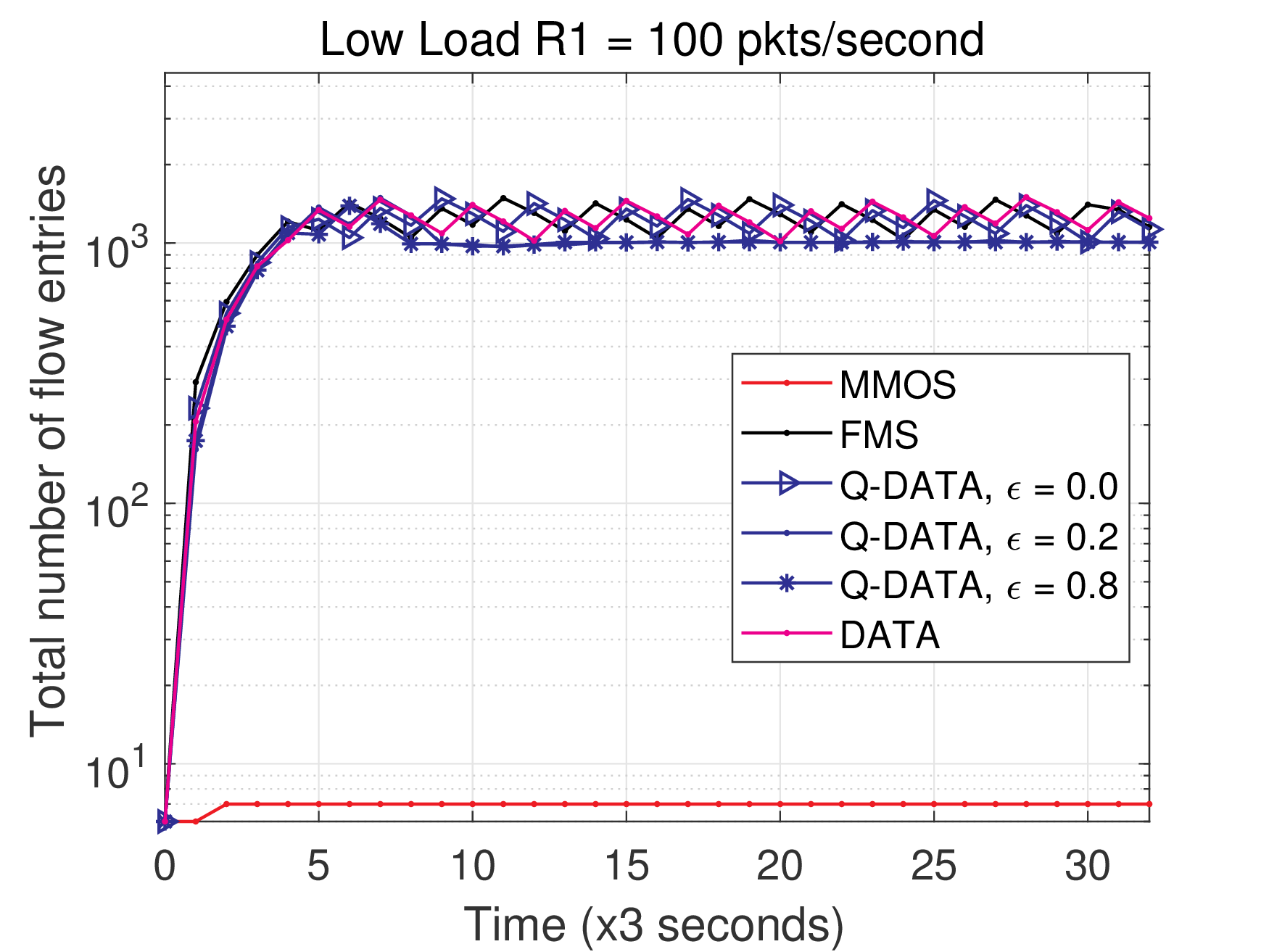}\hfill
  }
  \subfloat[]{
  \includegraphics[width=.3\linewidth]{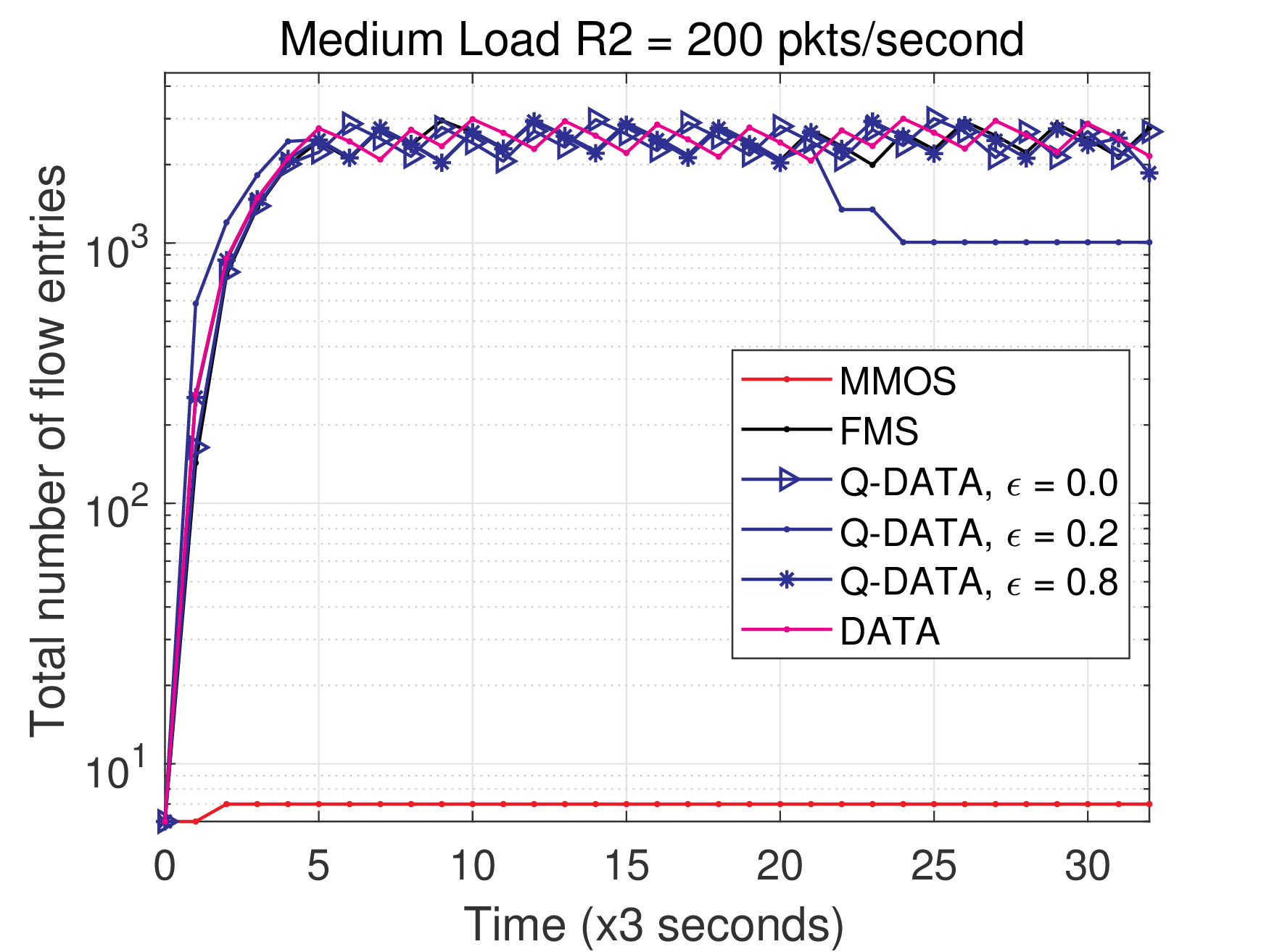}%
  }
  \subfloat[]{
  \includegraphics[width=.3\linewidth]{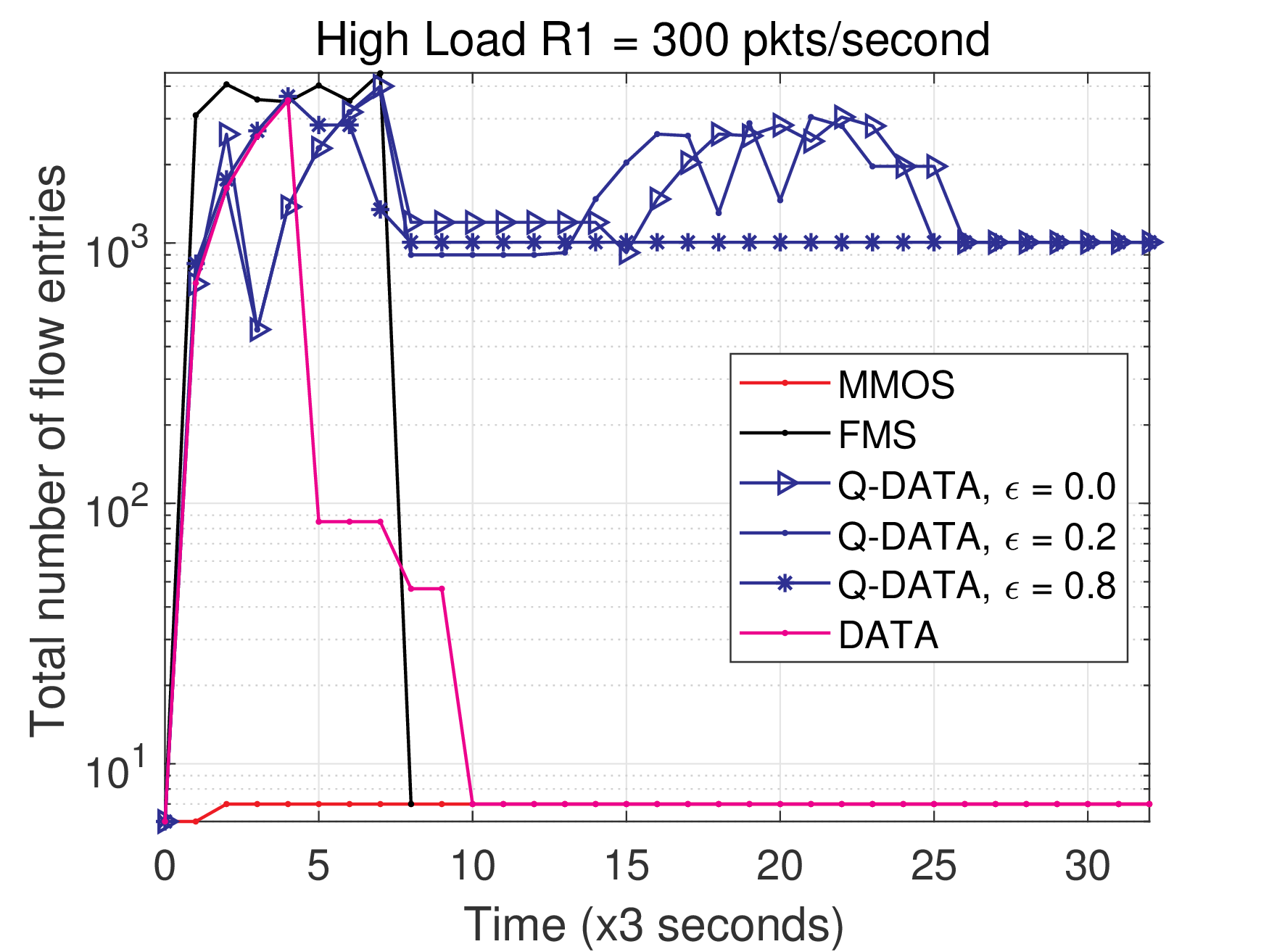}%
  }
  \end{minipage}%
  \caption{Total number of flow entries in the SDN switch for three different traffic loads: (a) Low Load $R1$=100, (b) Medium Load $R2$=200 and (c) High Load $R3$=300}
  \label{fig:FlowNumber}
\end{figure*}

As can be seen in Fig. \ref{fig:FlowNumber}, the MMOS scheme accounts for a very low amount of traffic flow entries in all scenarios. For the low and medium load cases, FMS, $Q$-DATA ($\epsilon=0.0$, $\epsilon=0.2$ and $\epsilon=0.8$) and DATA are supposed to have the same amount of traffic flow entries in the switch since the total number of flow entries is always below the critical level (${f_{cap}}_{i}$). Note, that there are some minor variations for $Q$-DATA with $\epsilon=0.2$ and $\epsilon=0.8$ because the $Q$-learning based Traffic Flow Matching Policy Creation module is allowed to take random actions that leads to a \textit{Good} state of the SDN switch with a high immediate reward value (average number of match fields of a flow entry) and to no further flow entry changes in the remaining time.

In the high load scenario, the FMS scheme leads to errors and exceptions after a short period of time causing a massive reduction in the number of flow entries because the SDN switch and the ONOS controller suspend their operation. In case of DATA, after reaching the switch's flow-table entry upper limit (${f_{cap}}_{i}$), the flow matching scheme is changed to MMOS for some destination hosts leading to a very small amount of flow entries in the remaining time. In contrast, the $Q$-DATA framework maintains a significant number of flow rules by applying appropriate traffic flow matching policies, e.g., a layer 2 \& layer 3 matching scheme which provides a higher traffic flow matching granularity and avoids the performance degradation of the switch. Besides, the $Q$-learning based Traffic Flow Matching Policy Creation module depends on future states, i.e., $s_{t+1}$, and tries to maximize the traffic flow matching granularity by changing to other schemes which provide more traffic flow information details. Therefore we observe some changes in the number of flow entries during our experiments.

\begin{figure*}
  \begin{minipage}{\linewidth}
  \centering
  \subfloat[]{
  \includegraphics[width=.3\linewidth]{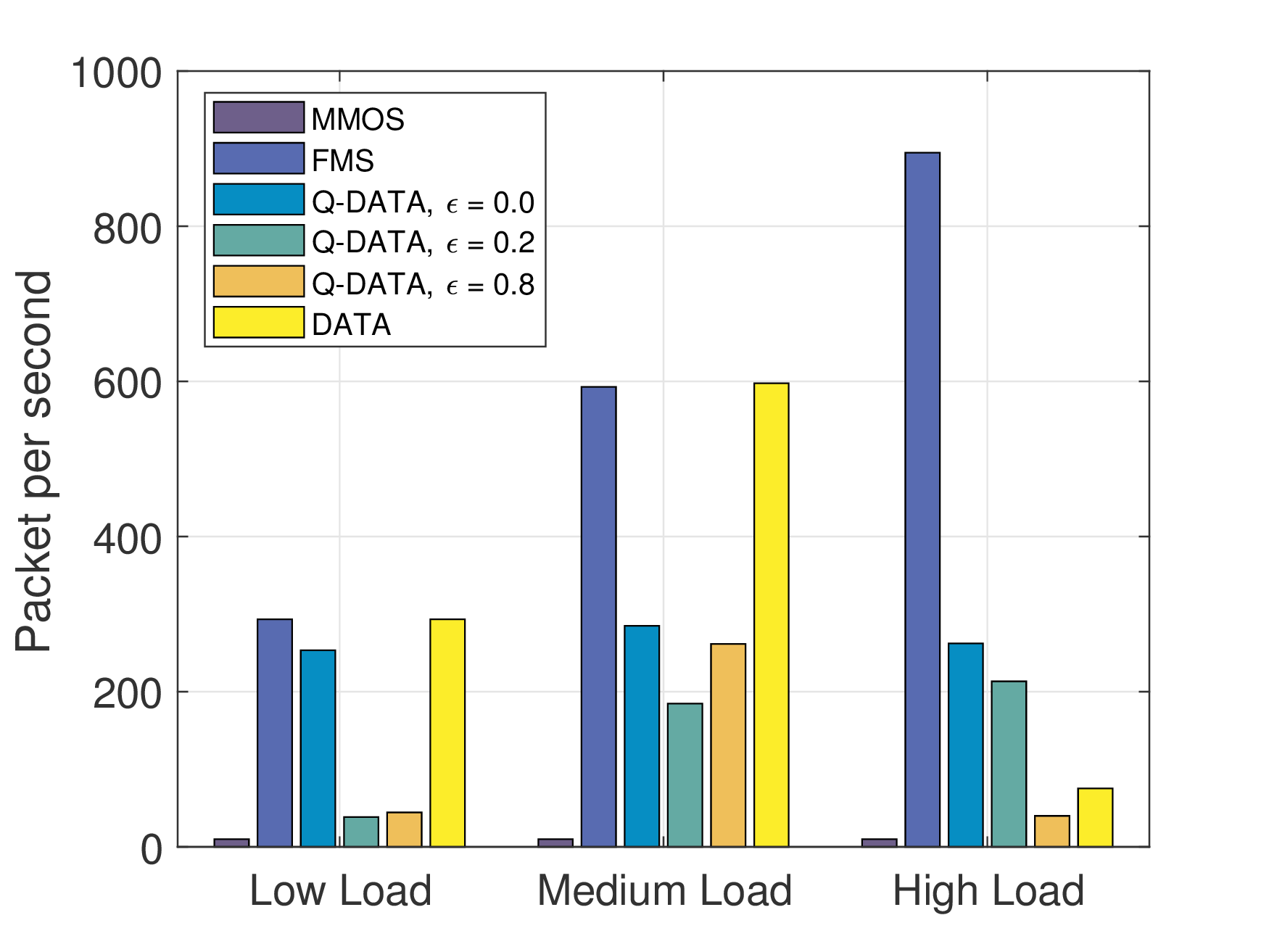}\hfill
  }
  \subfloat[]{
  \includegraphics[width=.3\linewidth]{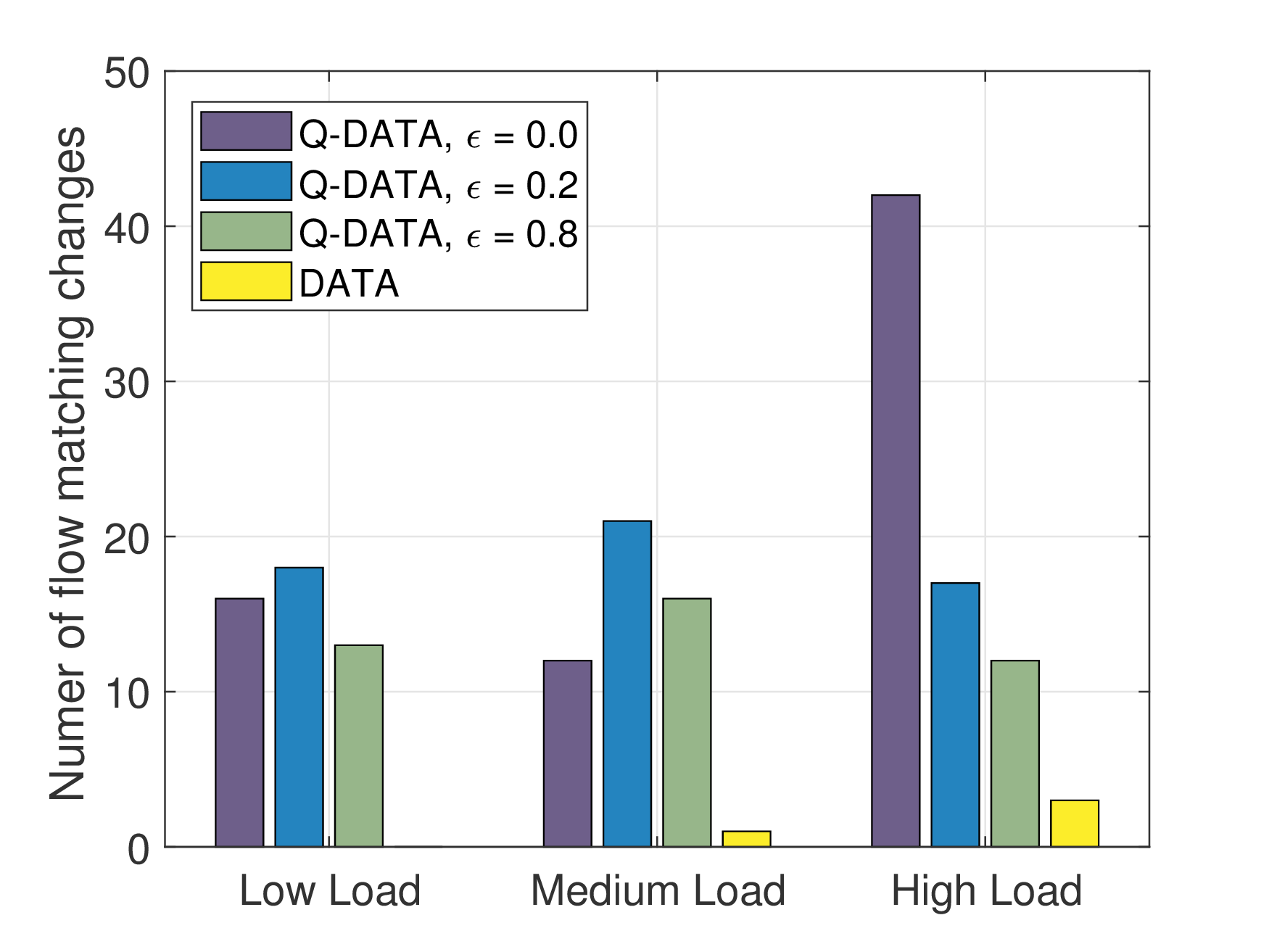}%
  }
  \subfloat[]{
  \includegraphics[width=.3\linewidth]{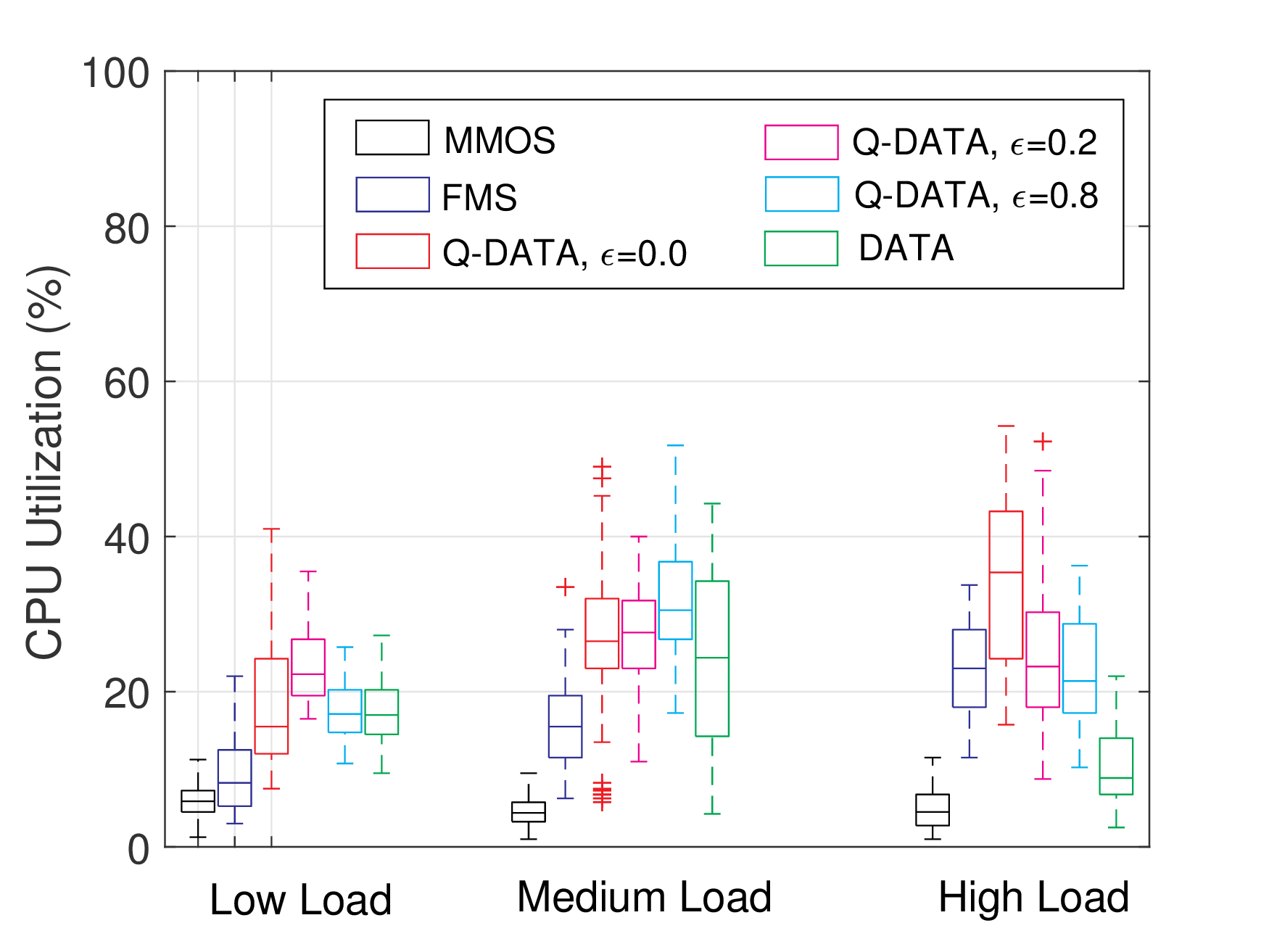}%
  }
  \end{minipage}%
  \caption{(a) Average \textit{packet\_in} rate (pkts/s) to the ONOS controller for different traffic flow matching schemes and traffic loads, (b) Total number of traffic flow matching scheme changes (during 500 seconds experiment duration) for different loads, (c) Average CPU utilization of the controller machine (during 500 seconds experiment duration)}
  \label{fig:PacketIn_Frequencey_CPU}
\end{figure*}

\subsubsection{Average \textit{packet\_in} message rate to the ONOS controller}
Fig. \ref{fig:PacketIn_Frequencey_CPU} (a) illustrates the average number of \textit{packet\_in} messages per second arriving at the Built-in Forwarding application. Contrary to the FMS and DATA schemes, for all traffic loads, the $Q$-DATA framework with the optimal traffic flow matching policy ($\epsilon=0.0$) allows the ONOS controller to process an acceptable \textit{packet\_in} rate. This significantly reduces the workload of the Built-in Forwarding application because of less new flow installation queries. The results for the $Q$-DATA scheme with $\epsilon=0.2$ and $\epsilon=0.8$ are expected to be better for a longer experiment duration.

\subsubsection{Errors and exceptions}
Another key criterion for the performance evaluation of the $Q$-DATA solution is the time until an error or exception (observed by the ONOS terminal) occurs due to a degraded SDN switch. Our measurements show that the FMS scheme causes \textit{disconnected} channels errors and \textit{FlowRuleManager} exceptions in the ONOS controller after 7 to 10 seconds since the high traffic load is generated. For the other traffic load cases, no errors and exceptions are observed.

\subsubsection{Frequency of changing flow matching policy}
We record the total number of traffic flow matching scheme changes of the proposed $Q$-DATA framework and the DATA scheme. As shown in Fig. \ref{fig:PacketIn_Frequencey_CPU} (b), the DATA scheme tries to keep the SDN switch in a \textit{Good} state as long as possible---therefore no changes in the traffic flow matching scheme occur for low and medium load scenarios, but some changes happen in the high load case (i.e., a change from FMS to MMOS). Contrary, $Q$-DATA performs some changes depending on newly incoming traffic flows in the switch. In particular, in the high load case, $Q$-DATA with $\epsilon=0.0$ performs several flow matching scheme changes, e.g., between layer 2 \& layer 3 matching and FMS, to provide more traffic flow information details while guaranteeing that the SDN switch forwarding performance does not degrade.

\subsubsection{Computational overhead}
Fig. \ref{fig:PacketIn_Frequencey_CPU} (c) shows measurements of the CPU utilization of the controller machine. It can be seen that the three $Q$-DATA scheme variants ($\epsilon=0.0$, $\epsilon=0.2$, $\epsilon=0.8$) consume more CPU resources for all traffic loads. This is due to the fact that the $Q$-DATA App actively monitors and analyzes the network traffic, especially in the case of high traffic load. It tries to maximize the traffic flow matching granularity and to avoid any performance degradation of the switch. Nonetheless, this computational overhead is acceptable considering the benefits of the $Q$-DATA Scheme.

\begin{table}
\centering
\caption{Anomaly detection performance of different traffic flow matching solutions and traffic loads}
\label{tab:SecurityAnalysis}       
\begin{tabular}{|c|c|c|c|c|c|c|}
\hline
\multicolumn{7}{|c|}{\textbf{TCP SYN flood attack detection performance (\%)}}\\ 
\hline
& MMOS & FMS & $Q$-DATA & $Q$-DATA & $Q$-DATA & DATA\\
&  &  & $\epsilon=0.0$ & $\epsilon=0.2$ & $\epsilon=0.8$ & \\
\hline
$R1$ & 0.0 & 98.02 & 97.05 & 96.04 & 96.53 & 97.52\\
\hline
$R2$ & 0.0 & 97.64 & 96.53 & 97.37 & 96.06 & 98.05\\
\hline
$R3$ & 0.0 & 0.0 & 85.20 & 81.20 & 82.00 & 0.0\\
\hline

\multicolumn{7}{|c|}{\textbf{Port scanning attack detection performance (\%)}}\\ 
\hline
& MMOS & FMS & $Q$-DATA & $Q$-DATA & $Q$-DATA & DATA\\
&  &  & $\epsilon=0.0$ & $\epsilon=0.2$ & $\epsilon=0.8$ & \\
\hline
$R1$ & 0.0 & 96.53 & 97.45 & 96.00 & 96.33 & 97.30\\
\hline
$R2$ & 0.0 & 98.22 & 96.43 & 97.01 & 96.62 & 97.22\\
\hline
$R3$ & 0.0 & 0.0 & 84.34 & 84.10 & 82.56 & 0.0\\
\hline

\multicolumn{7}{|c|}{\textbf{Low and Slow DoS attack detection performance (\%)}}\\ 
\hline
& MMOS & FMS & $Q$-DATA & $Q$-DATA & $Q$-DATA & DATA\\
&  &  & $\epsilon=0.0$ & $\epsilon=0.2$ & $\epsilon=0.8$ & \\
\hline
$R1$ & 0.0 & 96.32 & 96.25 & 95.70 & 97.12 & 95.78\\
\hline
$R2$ & 0.0 & 96.34 & 97.36 & 96.54 & 96.21 & 95.92\\
\hline
$R3$ & 0.0 & 0.0 & 88.24 & 85.32 & 85.45 & 0.0\\
\hline
\end{tabular}
\end{table}

\subsection{Anomaly Detection Performance Results}
In order to show the enhancement of the traffic flow monitoring capability provided by the $Q$-DATA framework, we evaluate the anomaly detection performance of the SOM-based IDS application for three attack types, i.e., TCP SYN flood\footnote{Attackers try to send as fast as possible TCP segments with different spoofed source IP addresses and TCP ports to the Web servers leading to a large number of new flow entries in the SDN switch in a short time period.}, Port scanning\footnote{Attackers try to send as many as possible TCP segments with different destination ports to the Web servers and wait for response packets.} and Low and Slow Denial-of-Service\footnote{Attackers periodically send requests as slow as possible with little resources and try to keep all installed flow entries in the SDN switch alive as long as possible, which renders the victim inaccessible.}. For the evaluation we apply the following fitness function: 
\begin{equation}\label{FitnessFunction}
F_{anomaly} = W_{D_{r}}D_{r} + W_{A_{c}}A_{c} + W_{F_{a}}e^{-F_{a}},
\end{equation}
where $D_{r}$ represents the Detection rate, $A_{c}$ Accuracy, $F_{a}$ False alarm rate and $W_{D_{r}}, W_{A_{c}}$ and $W_{F_{a}}$ are weight values which are equally set to 1/3 in our evaluation. 

As shown in Table \ref{tab:SecurityAnalysis}, no alert is raised in case of the MMOS scheme for all attack types and traffic loads because the traffic towards the Web servers is grouped into flow entries in the switch that makes it for the IDS impossible to detect any attacks. In the low and medium load scenarios, for the FMS, $Q$-DATA ($\epsilon=0.0$, $\epsilon=0.2$, $\epsilon=0.8$) and DATA schemes, three attacks are detected by the IDS with similar levels of attack detection performance.

In the high traffic load case, for FMS, the operation of the SDN controller and the switch are suspended. This makes the IDS application unable to gather traffic information from the SDN controller and to detect the attacks. For the DATA scheme the SDN switch stays operational, however traffic flows targeting to the servers are aggregated to some MMOS flows in the SDN switch. Hence there is no chance\footnote{Nevertheless, for a larger scale network that comprises several switches (like the enterprise network in our previous study \cite{DATA}), the SOM-based IDS is expected to achieve a good attack detection performance as some switches carry attack traffic flows and still stay operational (i.e., are in a \textit{Good} forwarding performance state).} to recognize malicious traffic flows towards the Web servers for all three attack types. Contrary, $Q$-DATA, by frequently changing between different flow matching schemes, provides more detailed traffic flow information and enables the IDS application to recognize the attack presence. However, because of the variation of statistics information caused by the traffic flow matching scheme change in the switch, the attack detection performance in case of high traffic load is lower than for low and medium traffic loads.

\section{Conclusion}\label{Conclution}
In this paper, we present a traffic flow matching control framework based on reinforcement learning called $Q$-DATA which improves traffic flow monitoring in SDN based networks and proactively prevents performance degradation of SDN switches. We conduct a comprehensive performance analysis of the $Q$-DATA framework. Our results show that---compared to the default behavior of common SDN controllers and to our previous DATA scheme---the new $Q$-DATA framework by applying always the optimal traffic flow matching policy yields remarkable performance benefits. In our future work, we intend to focus on an optimized integration of traffic flow matching control and traffic anomaly detection. 

\section{Acknowledgments}
This work has been performed in the framework of the Celtic-Plus project SENDATE Secure-DCI, funded by the German BMBF (ID 16KIS0481).

\appendices
\section{Algorithms from our previous work \cite{DATA}}

\begin{algorithm}[H]
\caption{Identification of the destination hosts whose flows are most critical to the performance of SDN switch $i$}
\label{FMStoMMOS}
\begin{algorithmic}
\STATE \textbf{Input}: $S_i$ = $\left\{(host_{1}, flow_{1}), (host_{2}, flow_{2}), ..., (host_{\chi}, flow_{\chi})\right\}$: set of destination hosts and respective number of flow entries associated with these hosts in switch $i$; $f_i$ = $\sum_{c=1}^{\chi}$ $flow_{c}$: total number of current flow entries in switch $i$; $index$ = 1: first index. \textbf{Output}: $H_{i}$ = $\left\{ \right\}$: set of destination hosts.
\STATE \textbf{begin}
\STATE Sort $S_i$ in descending order of the current flow $flow_c$ (from highest to lowest numbers).
\LOOP
\STATE $H_i$.$append$($S_{i}$[$index$])
\STATE $f_{remaining}$ = 1 + $\sum_{c=index+1}^{\chi}flow_{c}$ \COMMENT{One MMOS flow entry is installed in switch $i$}.
\STATE $\Delta f_{i}$ = $f_i$-$f_{remaining}$ \COMMENT{Delete $\Delta f$ flow entries in switch $i$}.
\STATE $sign$ = $SVM(f_{remaining},\Delta f_{i})$
	\IF {$sign$ = +1}
		\STATE \textbf{break}
        \COMMENT{Switch $i$ can handle $f_{remaining}$ entries}.
	\ELSE
    	\STATE $index$ = $index+1$
    	\COMMENT{Switch $i$ cannot handle $f_{remaining}$ entries}.
	\ENDIF
\ENDLOOP
\STATE \textbf{return} $H_i$.
\end{algorithmic} 
\end{algorithm}

\begin{algorithm}[H]
\caption{Identification of the MMOS flows/destination hosts related to SDN switch $i$ for which changing back to FMS is feasible}
\label{MMOStoFMS}
\begin{algorithmic}
\STATE \textbf{Input}: $S_i$ = $\left\{(host_{1}, R_{pkt_1}), (host_{2}, R_{pkt_2}), ..., (host_{\chi}, R_{pkt_{\chi}})\right\}$: set of destination hosts and respective packet rate of MMOS flows associated with these hosts in switch $i$; ($f_i$, $f_{cap_{i}}$): total number of current flow entries and maximum number of flow entries in switch $i$; $f_{extra}$: number of flow entries that might be added in switch $i$. \textbf{Output}: $H_{i}$ = $\left\{ \right\}$: set of destination hosts.
\STATE \textbf{begin}
\FOR{$index = 1$; $index \leq z$; $index$++}
    \STATE $f_{extra}$ = $idle\_timeout$*$R_{pkt_{index}}$ \COMMENT{Worst case assumption: each packet is associated with a new entry in switch $i$}.
  \IF {($f_{extra}$ + $f_i$ $)<$ $f_{cap_{i}}$}
    \STATE $H_{i}$.append[$h_{index}$].
  \ELSE
    \STATE \textbf{continue}
  \ENDIF
\ENDFOR
\STATE \textbf{return} $H_{i}$.
\end{algorithmic} 
\end{algorithm}

\bibliographystyle{ieeetr}
\bibliography{References.bib}

\end{document}